\documentclass[10pt]{iopart}
\usepackage{graphicx,epsfig}

\begin{document}
\title{Response of an atomic Bose-Einstein condensate to a rotating elliptical trap}
\author{N.G. Parker and C.S. Adams}
\address{Department of Physics, University of Durham, South Road,
Durham DH1 3LE, United Kingdom}

\begin{abstract}
We investigate numerically the response of an atomic Bose-Einstein
condensate to a weakly-elliptical rotating trap over a large range
of rotation frequencies.  We analyse the quadrupolar shape
oscillation excited by rotation, and discriminate between its
stable and unstable regimes. In the latter case, where a vortex
lattice forms, we compare with experimental observations and find
good agreement. By examining the role of thermal atoms in the
process, we infer that the process is temperature-independent, and
show how terminating the rotation gives control over the number of
vortices in the lattice. We also study the case of critical
rotation at the trap frequency, and observe large centre-of-mass
oscillations of the condensate.
\end{abstract}
\pacs{03.75.Lm,03.75.Kk}

\section{Introduction}
Recent experiments on dilute atomic Bose-Einstein condensates
(BECs) have generated vortex lattices by mechanically rotating the
system \cite{madison2000,hodby,abo-shaeer2001}, in analogy to the
classic rotating bucket experiment in liquid Helium
\cite{osborne}. The smooth magnetic trap confining the condensate,
with trap frequency $\omega_r$, is deformed anisotropically and
rotated at frequency $\Omega$. Shape oscillations of the
condensate are excited, and the quadrupole mode plays a pivotal
role in the evolution of the system: for low rotation frequencies,
$\Omega<0.7\omega_r$, a long-lived quadrupole mode has been
observed \cite{quad_expts}, while for higher rotation frequencies
$\Omega \sim 0.7\omega_r$, vortex lattice formation has been seen
\cite{madison2000,hodby,abo-shaeer2001}.  In the latter case,
vortex formation is seeded by a dynamical instability of the
quadrupole mode \cite{recati,sinha,madison2001}. Experimental
results suggest that the crystallisation of the vortices is
temperature-independent \cite{abo-shaeer2002}. Indeed, the full
process of lattice formation has been successfully simulated using
the zero temperature Gross-Pitaevskii equation
\cite{lundh,lobo,parker_lattice} (although other studies found it
necessary to include damping effects \cite{tsubota,penckwitt}).
These results have also indicated that lattice formation is a
two-dimensional effect.  An experimental aim in this area is to
examine the regime of fastly rotating BECs, and probe exotic
phenomena such as vortex melting and the bosonic quantum Hall
effect \cite{quantum_hall}. This has led to the experimental
investigation of a BEC under critical rotation \cite{crit_rot},
where the rotation frequency equals the trap frequency, and
super-critical rotation \cite{fast_rot}.

In this work we investigate the intrinsic reponse of an atomic BEC
to rotation in a weakly-elliptical trap using the zero-temperature
Gross-Pitaevskii equation. This is performed over a large range of
rotation frequencies, allowing us to probe diverse regimes of
condensate dynamics relevant to current BEC experiments. In the
regime of sub-critical rotation, a quadrupole shape mode dominates
the dynamics.  Close to resonance this mode is unstable and vortex
lattice formation occurs, in close agreement with experiments. We
simulate the role of thermal atoms in the lattice formation, and
consider the effect of terminating the rotation.  At critical
rotation, $\Omega \approx \omega_r$, we show that large
centre-of-mass oscillations dominate the dynamics.

At ultra-cold temperature $T\ll T_{\rm c}$ (where $T_{\rm c}$ is
the BEC transition temperature) and for a large number of
condensate atoms, thermal and quantum effects are minimal, and an
atomic BEC can be approximated by a mean-field `macroscopic
wavefunction' $\psi({\bf r},t)$. The dynamics of this wavefunction
are well-described by the Gross-Pitaevskii equation (GPE)
\cite{BEC},
\begin{equation}
i\frac{\partial \psi({\bf r},t)}{\partial
t}=\left[-\frac{\hbar^2}{2m}\nabla^2+V_{\rm ext}({\bf
r},t)+g|\psi({\bf r},t)|^2\right]\psi({\bf r},t),
\end{equation}
where $V_{\rm ext}({\bf r},t)$ is the external confinement and $m$
is the atomic mass.  The nonlinear coefficient $g=4\pi \hbar^2
Na/m$ governs the atomic interactions, where $N$ is the number of
atoms and $a$ is the {\it s}-wave scattering length.  It is useful
to consider the hydrodynamic interpretation of the wavefunction
$\psi({\bf r},t)=\sqrt{n({\bf r},t)}\exp[i\phi({\bf r},t)]$, where
$n({\bf r},t)$ is the atomic density and $\phi({\bf r},t)$ is a
phase factor, which defines a fluid velocity via $v({\bf
r},t)=(\hbar/m)\nabla \phi({\bf r},t)$.

Vortices in BEC are characterised density node and an azimuthal
phase slip, quantized to multiples of $2\pi$ (to keep $\psi$
single-valued). The lowest energy state of a BEC in a rotating
system (above the critical rotation for the onset of vortices) is
an ordered triangular lattice of singly-charged vortices
\cite{fetter} (vortices with multiple `charge' are unstable
\cite{shin}), whereby the system mimics solid-body rotation.
However, an energy barrier exists for vortices to enter the
condensate and so the transformation process from a vortex-free
state to a vortex lattice is non-trivial. Lattices have been
formed thermodynamically \cite{penckwitt,williams} by condensing a
rotating thermal cloud \cite{haljan}. In \cite{raman2}, the
precession of a small laser `stirrer' above a critical velocity
generated vortex excitations locally \cite{raman1} and ultimately
a lattice \cite{lundh}. Other lattice experiments
\cite{madison2000,hodby,abo-shaeer2001} involve the direct
rotation of the condensate in an anisotropic trap, which is the
case we will study in this paper.  In addition we consider a
sudden turn on of the trap rotation rather than the adiabatic
introduction of the rotating anisotropy \cite{sinha,madison2001}.
Condensate shape oscillations are excited, and through a dynamical
instability of the dominant quadrupole mode
\cite{recati,sinha,madison2001}, the condensate fragments
\cite{parker_lattice}.  Provided the two-fold rotational symmetry
of the system is allowed to break, a turbulent state of sound
waves (density excitations) and vortices is generated.
Subsequently, through the interaction of the vortices with the
sound field \cite{vortexsound}, the vortices gradually become
ordered into a lattice configuration.

The trap rotation, which occurs in the {\it x-y} plane about the
{\it z}-axis, strongly polarizes the system. Although bending and
Kelvin wave excitation of vortex lines do occur \cite{bending},
the 2D dynamics in the rotating plane dominate, and a 2D
description is sufficient to simulate lattice formation
\cite{lundh,parker_lattice,tsubota}. We therefore proceed by
solving the computationally-advantageous 2D GPE in the rotating
{\it x-y} plane.

In the experiments of \cite{madison2000,hodby,madison2001} the
trap is approximately elliptical in the rotating plane with the
form,
\begin{eqnarray}
V_{\rm ext}(x,y,t)=\frac{1}{2}m\omega_r^2
\left[(x^2+y^2)+\varepsilon (X(t)^2-Y(t)^2)\right]. \label{eqn:V}
\end{eqnarray}
Here the first term represents the static radial harmonic trap
with frequency $\omega_r$. In the second term, $\varepsilon$ is
the ellipticity parameter of the trap (typically $\varepsilon\ll
1$), and the rotating coordinates $X(t)$ and $Y(t)$ are given by,
\begin{eqnarray}
X(t)&=&x\cos(\Omega t)-y\sin(\Omega t) \\
Y(t)&=&x\sin(\Omega t)+y\cos(\Omega t),
\end{eqnarray}
where $\Omega$ is the trap rotation frequency.  Note that in
\cite{abo-shaeer2001} the trap is strongly anharmonic, rather than
weakly-elliptical.

Our units of length and time are the healing length
$\xi=\hbar/\sqrt{m\mu}$ and $(\xi/c)$, where $c=\sqrt{\mu/m}$ is
the Bogoliubov speed of sound. The chemical potential $\mu=n_0g$
is the unit of energy, where $n_0$ is the peak atomic density.
Unless stated otherwise, we employ parameters which match the
experiment of Madison {\it et al.} \cite{madison2000}: our trap
ellipticity is $\varepsilon=0.025$, the chemical potential is
$\mu=10\hbar \omega$, the trap frequency is $\omega_r=0.1
(c/\xi)$, and the Thomas-Fermi radius of the condensate is $R_{\rm
TF}=14\xi$. For such a $^{87}$Rb condensate, our units of time and
space correspond to $\xi\sim 0.2~\mu$m and $(\xi/c)\sim
10^{-4}~$s.

Our initial state is found by propagating the GPE in imaginary
time ($t\rightarrow -it$) subject to the initial potential $V_{\rm
ext}(x,y,t=0)$ while forcing the peak density to unity.
Subsequently the system is run in real time.  We generally
characterise the state of the system in terms of the condensate
energy, which is evaluated according to the GPE energy functional,
\begin{equation}
E=\int \left[-\frac{\hbar^2}{2m}|\nabla \psi|^2+V_{\rm
ext}|\psi|^2+\frac{g}{2}|\psi|^4  \right] {\rm d}x {\rm d}y.
\end{equation}
Note that we subtract from equation (5) the energy corresponding
to the initial state, i.e. energy is measured relative to the
ground state.

In section \ref{sec:quad} we investigate the quadrupole mode of a
condensate in a rotating trap, and discriminate between the stable
and unstable regimes.  In the latter case, we observe vortex
lattice formation, and in section \ref{sec:expts} we compare our
results with the lattice experiments. Section \ref{sec:thermal}
infers the role of thermal atoms in the lattice formation, and the
effect of terminating the rotation is analysed in section
\ref{sec:term}. The condensate dynamics under critical and
super-critical rotation of the trap are studied in section
\ref{sec:crit}. Finally, section \ref{sec:conc} presents the
conclusions of this work.

\section{The Quadrupole Mode}
\label{sec:quad}

In the rotating frame, and for slow rotations, the stationary
solution in an elliptical trap is an elliptical condensate
\cite{recati,feder}, with the size of the ellipticity depending
not only on the trap ellipticity $\varepsilon$ but also the
rotation frequency $\Omega$. Such a state can be reached by
adiabatically ramping up the rotation frequency from $0$ to the
chosen value, $\Omega$ \cite{sinha}. This technique has been
implemented experimentally \cite{madison2001}.  In our case,
however, the rotation is suddenly turned on with frequency
$\Omega$ and maintained at this value.  The condensate is
continuously driven out of equilibrium and never reaches a steady
elliptical state. Consequently, collective shape excitations are
expected to occur.

Classical hydrodynamics predict that, for a non-rotating trapped
condensate, the shape oscillations satisfy the dispersion relation
$\omega_\ell=\sqrt{\ell}\omega_r$, where $\ell$ is the angular
momentum quantum number \cite{stringari}.  The two-fold symmetry
of our system implies that only even angular momentum modes will
be excited, and we expect the lowest order mode to dominate, i.e.
the $\ell=2$ quadrupolar mode with frequency $\omega^0_{\rm
Q}=\sqrt{2}\omega_r$.  When driven resonantly at half this
frequency, i.e. at $\Omega=\omega_r/\sqrt{2}$, this mode is
predicted to be dynamically unstable \cite{sinha}, leading to
vortex lattice formation. Experimental results
\cite{madison2000,hodby,abo-shaeer2001} and simulations of the GPE
\cite{lundh,lobo,parker_lattice,tsubota} have verified this
instability, and that it occurs over a range of rotation
frequencies in the vicinity of $\omega_{\rm Q}^0$.

\subsection{Stable quadrupole mode}

\begin{figure}[b]
\begin{center}
\includegraphics[width=13cm,angle=0]{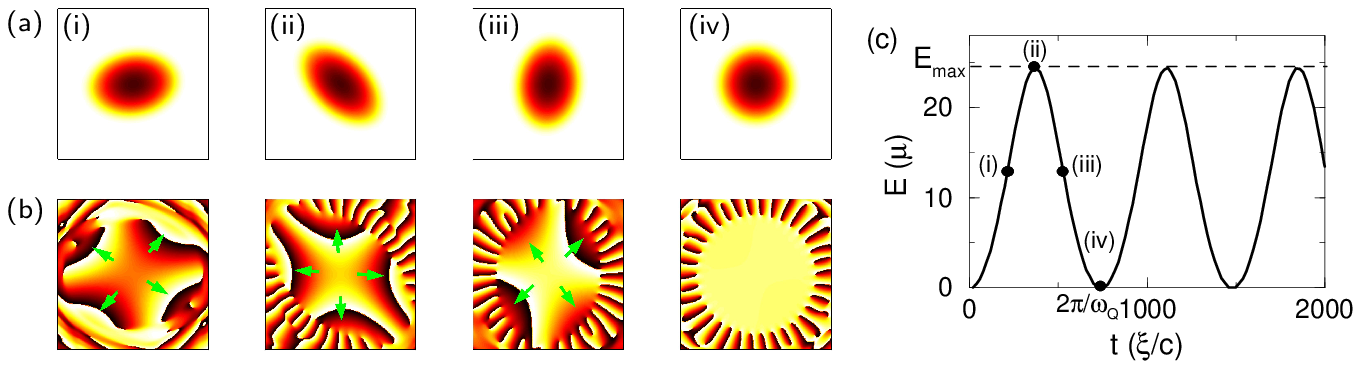}
\caption{Stable quadrupole mode. (a)-(b) Snapshots in density and
phase during one oscillation of quadrupolar flow
($\Omega=0.7\omega_r$), at times $t=$ (i) $185$, (ii) $370$, (iii)
$555$, and (iv) $740~(\xi/c)$. Each box corresponds to the region
$[-25,25]\xi\times[-25,25]\xi$. Arrows in (i)-(iii) indicated the
direction of fluid flow.  Density and phase are shown in the
ranges $(0-n_0)$ and $(-\pi,\pi)$ respectively, with light/dark
regions corresponding to low/high values. Note that the complex
phase structure at large radii has no meaning since the density is
essentially zero. (c) Energy of the condensate during quadrupolar
flow. The times of the snapshots in (a)-(b) are indicated. }
\label{stable_quad}
\end{center}
\end{figure}

We first consider the case of the stable quadrupolar mode.  Figure
\ref{stable_quad}(a)-(b) shows the condensate density and phase
throughout a cycle of the quadrupole oscillation for a rotation
frequency $\Omega=0.7\omega_r$. Note that, due to the rotating
trap, the two primary axes of the quadrupole continuously rotate
over time at the rotation frequency $\Omega$.

The condensate density initially has a circular shape. Over time
it becomes elongated (figure \ref{stable_quad} (a)(i)) up to some
maximum point (figure \ref{stable_quad}(a)(ii)). The elongation
then decreases (figure \ref{stable_quad}(a)(iii)) and the
condensate returns to a circular state (figure
\ref{stable_quad}(a)(iv)). The phase distribution initially, and
after one complete cycle, is uniform. As the condensate becomes
elongated (figure(b)(i)-(iii)), the phase develops a quadrupolar
structure of the form $\phi(x,y)\propto XY$, where $X$ and $Y$ are
the rotating coordinates of the ellipse. This corresponds to a
flow of fluid towards and away from the trap centre in different
quadrants of the ellipse, as indicated by the arrows in figure
\ref{stable_quad}(b)(i)-(iii). Although the condensate {\it
appears} to be rotating, the combined effect of the quadrupole
cycle and the rotation of the quadrupole axes ensures that the
fluid remains irrotational at all times \cite{quad_expts,feder}.
When the condensate is highly elongated (figure
\ref{stable_quad}(ii)), the phase structure is dense, i.e. the
fluid velocity is high, as required to maintain the irrotational
flow.

The condensate energy (figure \ref{stable_quad}(c))  oscillates
sinusoidally at the quadrupole frequency $\omega_{\rm Q}$ between
$E=0$, representing the circular, ground state condensate, and a
maximum value $E=E_{\rm max}$, corresponding to the fully
elongated, excited state.  This is analogous to a vortex cycling
in and out of a condensate driven by a narrow optical stirrer
\cite{caradoc}, and to the Rabi oscillations of an atom in a light
field.  Note that the energy shows no additional behaviour: we do
not detect higher order modes (e.g. $\ell=4, 6,...$), and the
orientation of the quadrupole (oscillating at $\Omega$) does not
affect the energy.

\subsection{Behaviour of the quadrupole mode with rotation frequency}

\begin{figure}[t]
\begin{center}
\includegraphics[width=4.0cm,angle=-90]{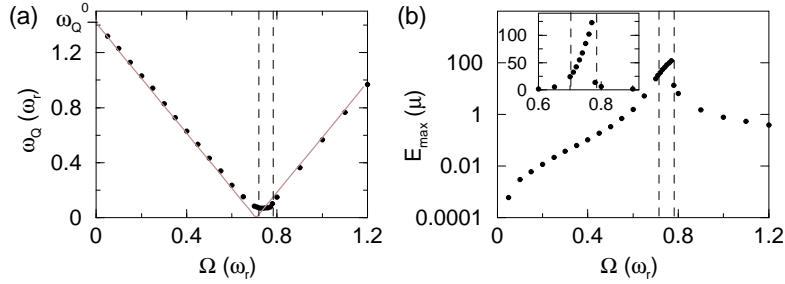}
\caption{The quadrupole mode. (a) Quadrupole frequency
$\omega_{\rm Q}$ as a function of rotation frequency $\Omega$ for
the numerical data (circles) and the classical hydrodynamic
prediction of equation \ref{eqn:quad_freq} (grey/brown line). (b)
Energy amplitude $E_{\rm max}$ of the quadrupole oscillations as a
function of rotation frequency. The {\it y}-values are shown on a
logarithmic scale. Inset:  the central region of the main plot
shown with decimal scaling on the {\it y}-axis. In all plots the
dashed lines indicate the extent of the unstable quadrupole
region.} \label{quad_mode}
\end{center}
\end{figure}

The sinusoidal energy evolution in figure \ref{stable_quad}(c) is
a general feature of the quadrupole mode, characterised by the
quadrupole frequency $\omega_{\rm Q}$ and amplitude $E_{\rm max}$.
We now examine the variation of these quantities with rotation
frequency $\Omega$. First, consider the quadrupole frequency,
shown in figure \ref{quad_mode}(a). In the limit $\Omega
\rightarrow 0$ the quadrupole frequency approaches $\omega^0_{\rm
Q}\equiv \sqrt{2}\omega_r$, which corresponds to the analytic
prediction in a non-rotating system \cite{stringari}. For a system
rotating at frequency $\Omega$ the Hamiltonian features an
additional centrifugal term $-\Omega L_z$ \cite{fetter}, where
$L_z$ is the {\it z}-component of the angular momentum . This adds
a shift of $-\ell\Omega$ to the dispersion relation. For the
quadrupole $\ell=2$ mode this gives,
\begin{equation}
\omega_Q(\Omega)=|\omega_{\rm Q}^0-2\Omega|. \label{eqn:quad_freq}
\end{equation}
This equation is plotted in figure \ref{stable_quad}(a)
(grey/brown line) and fits the numerical data well for rotation
frequencies away from the quadrupole resonance at
$\Omega=\omega^0_{\rm Q}/2$. However, close to this resonance, the
data deviates from the linear trend and never reaches zero.  This
is the unstable region of the quadrupolar mode, which ultimately
leads to vortex lattice formation \cite{parker_lattice}. Note that
the unstable regime lies in the vicinity of, but not necessarily
centred around, the analytic prediction of $\Omega=\omega^0_{\rm
Q}/2$. This shift is due to the elliptical perturbation on the
system. The extent of this unstable region and how it compares to
experimentals will be discussed in section \ref{sec:expts}. The
amplitude of the quadrupole mode, shown in figure
\ref{quad_mode}(b), increases towards the region of instability.
Close to and within this region, the amplitude becomes extremely
large. The unstable quadrupole region is therefore associated with
when the quadrupole mode has a frequency close to zero and a large
amplitude.

\section{Comparison with experimental observations}
\label{sec:expts}

\subsection{Width of the unstable region and number of vortices}

\begin{figure}[h]
\begin{center}
\includegraphics[width=4.cm,angle=-90]{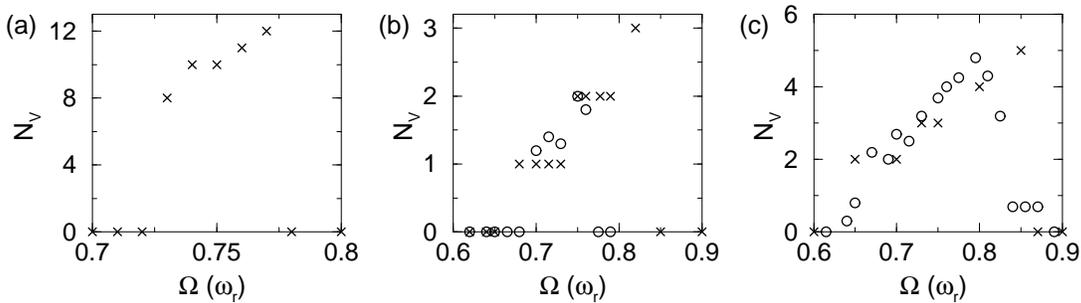}
\caption{Number of vortices in the final lattice versus rotation
frequency for (a) the system of Madison {\it et al.}
\cite{madison2000} with trap ellipticity $\varepsilon=0.025$, and
the system of Hodby {\it et al.} \cite{hodby} with trap
ellipticities of (b) $\varepsilon=0.041$ and (c)
$\varepsilon=0.084$. Numerical results are show by crosses. In
(b)-(c) the experimental data (averaged over four observations)
from \cite{hodby} is shown by circles.} \label{hodby}
\end{center}
\end{figure}

In the experiment of Madison {\it et al.} \cite{madison2000},
lattice formation was observed in the region
$0.68\leq\Omega/\omega_r\leq0.78$. Below this no vortices were
observed, while above this the condensate became turbulent. Figure
\ref{hodby}(a) shows the final number of vortices we observe in
the lattice for different rotation frequencies. To be consistent
with the experimental observations, where density imaging has
limited accuracy, we only count a vortex if it resides in a region
of significant density, which we take to be where the density is
greater than $n\approx 0.2n_0$. This also avoids the counting of
the phase singularities at very low density which have negligible
energy.  The range of rotation frequencies over which we observe
lattice formation, $0.72<\Omega/\omega_r<0.78$, is comparable to
the experimental observations. Within this region, we observe the
final number of vortices to increase with rotation frequency up to
12 vortices.

Both below {\it and above} the lattice-forming region we observe
stable quadrupolar oscillations, with no indication of a turbulent
regime.  Note that for super-critical rotations
$\Omega/\omega_r>1$ no more condensate atoms were observed in the
experiment. We will consider this regime in section
\ref{sec:crit}.

We now consider the experiment of Hodby {\it et al.} \cite{hodby},
and employ a system with chemical potential $\mu\approx4\hbar
\omega_r$ and trap frequency $\omega_r\approx 0.25 (c/\xi)$, which
has a Thomas-Fermi radius of $R_{\rm TF}\approx 6\xi$. Figure
\ref{hodby}(b)-(c) shows the number of vortices produced as a
function of $\Omega$ for two different trap ellipticities of (b)
$\varepsilon=0.041$ and (c) $\varepsilon=0.084$.  The experimental
observations of Hodby {\it et al.} \cite{hodby} are shown by
circles, while the numerical results from the GPE are shown by
crosses.  We see good agreement between theory and experiment,
both in the extent of the lattice-forming region and the number of
vortices produced, and particularly for the higher value of the
ellipticity. By comparing figures \ref{hodby}(b) and (c), we see
that the width of the unstable regime and the number of vortices
produced increase with trap ellipticity. The number of vortices
produced here is lower than for the parameters of Madison {\it et
al.} since the condensate is smaller. Note that Lundh {\it et al.}
\cite{lundh} have modelled this experiment using the 2D GPE
(albeit for different data) and also found good agreement.

\subsection{Symmetry breaking and the timescale for lattice formation}

In our simulations of the unstable quadrupole regime, numerical
noise grows exponentially and ultimately breaks the symmetry on a
macroscopic level \cite{parker_lattice}.  In real systems, noise
arises from many sources, e.g. thermal/quantum fluctuations, trap
imperfections and stray fields, and will have a similar
symmetry-breaking effect, albeit on a much larger scale. In order
to model such `real-life' effects we impose a random `jitter' on
the trap centre, limited to the region $[-\gamma,\gamma]\times
[-\gamma,\gamma]$. The parameter $\gamma$ therefore allows us to
control numerically the degree of symmetry-breaking. Conveniently,
even a very small jitter, e.g. $\gamma=0.0001\xi$, is sufficient
to dominate over the effects of numerical noise
\cite{parker_lattice}.  However, the final state of the lattice,
e.g. the number of vortices, is insensitive to $\gamma$. We will
employ a jitter of $\gamma=0.1\xi$, which corresponds to around
$20~$nm.

\begin{figure}[h]
\begin{center}
\includegraphics[width=4.5cm,angle=-90]{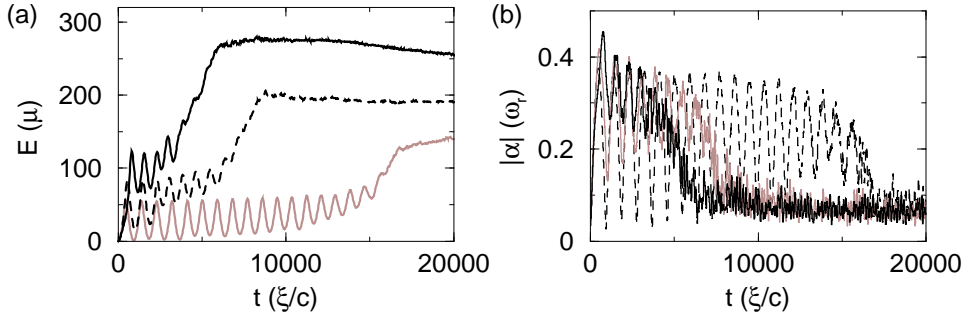}\\
\caption{(a) Evolution of the condensate energy for rotation
frequencies $0.73$ (solid grey/brown line), $0.75$ (dashed black
line) and $0.77$ (solid black line). (b) Evolution of the
elliptical deformation of the condensate, defined by equation
(\ref{eqn:alpha}), for the corresponding cases.  A trap jitter of
$\gamma=0.1\xi$ is employed. Note that the time unit corresponds
to $(\xi/c)\sim 10^{-4}~$s.} \label{madison}
\end{center}
\end{figure}

We return to using the system parameters of Madison {\it et al.}
\cite{madison2000}. Figure \ref{madison}(a) shows the evolution of
the condensate energy for various rotation frequencies within the
unstable region. Figure \ref{madison}(b) shows the corresponding
evolution of the elliptical deformation of the condensate $\alpha$
defined as,
\begin{equation}
\alpha=\Omega\frac{<X^2>-<Y^2>}{<X^2>+<Y^2>}, \label{eqn:alpha}
\end{equation}
where $<X^2>=\int X^2|\psi(x,y=0)|^2 {\rm d} X$ and $<Y^2>=\int
Y^2|\psi(x=0,y)|^2 {\rm d} Y$.  These curves have the same
qualitative features, which are related to the key stages in the
evolution \cite{parker_lattice}.
\begin{description}
    \item [Fragmentation:] The quadrupole mode becomes over-excited
    and atoms are ejected.  Partial quadrupole oscillations occur (where the condensate does not fully return to its circular
    state), with the energy undergoing partial oscillations.
    \item[Symmetry breaking and turbulence:] When the system asymmetry has
    grown to macroscopic levels the quadrupole oscillations
 cease. Energy is rapidly coupled into the condensate up to
some maximum value.  At this point the condensate is a turbulent
state of vortices and sound waves. Note it also returns to having
an approximately circular shape.
    \item[Crystallisation:] The total energy stays approximately
    constant, while vortex-sound interactions allow the vortices
    to settle into a lattice configuration.
\end{description}

As apparent in figure \ref{madison}, the length of the
fragmentation stage is highly sensitive to the trap rotation
frequency.  In the shortest case (solid black line in figure
\ref{madison}), the fragmentation stage persists for a time of
around $3000(\xi/c)$ (or $300~$ms).  On the other hand, the
timescale for the symmetry-breaking/turbulent stage is insensitive
to these effects, and typically extends for a time of around
$1000(\xi/c)$ (or $100~$ms). The crystallisation phase in our
simulations is also independent of these effects and remains a
slow process, typically taking $50000-10000~(\xi/c)$ (or
$0.5-1~$s) to form a well-ordered lattice.

In the experiments \cite{madison2000,hodby,madison2001}, the
condensate is observed to undergo partial quadrupole oscillations
(where the condensate does not fully return to a circular state),
become disordered, and finally form an ordered vortex lattice.
This appears to be consistent with the stages of lattice formation
described above. Indeed, in \cite{madison2001} the elliptical
deformation of the condensate is measured as a function of time,
and bears the same qualitative features as in figure
\ref{madison}(b).

We now compare our simulations to the experimentally-observed
timescales, where possible. In \cite{madison2001}, the partial
shape oscillations occur for approximately $250~$ms. Subsequently,
the experimental shape oscillations cease and the condensate
returns to an approximately circular shape. This occurs over a
time of around $50~$ms.  These timescales are comparable with the
quickest evolution in figure \ref{madison} (solid lines).

In the experiments, a lattice is observed after $(0.5-1)~$s of
trap rotation, where in our simulations \cite{parker_lattice} it
takes the equivalent of approximately 10 seconds for the lattice
to become well-ordered. We have shown that the timescale for the
fragmentation, symmetry-breaking and turbulent phases are
consistent with experimental observations, and so this time
anomaly must lie with the crystallisation phase. This discrepancy
in the crystallisation rate may be due to thermal dissipation
(although results in section \ref{sec:thermal} would contradict
this) or 3D effects such as Kelvin wave excitation and damping
\cite{bending}, both of which are not present in our simulations.

\section{The role of `thermal' atoms}
\label{sec:thermal}

In the classical field interpretation \cite{norrie05}, the GPE
describes both the mean-field condensate {\it and} the fluctuating
thermal field of random excitations. The thermal excitations
typically have high energy and momenta in comparison to the
condensate.  As described in \cite{parker_lattice}, an integral
part of the lattice formation process is the irreversible
generation of a highly-energetic outer cloud, which we will
interpret as thermal atoms.

\begin{figure}[h]
\begin{center}
\includegraphics[width=4.3cm,angle=-90]{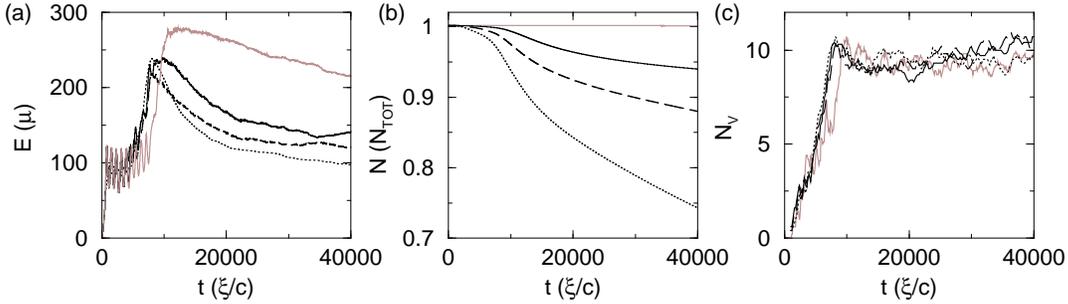}
\caption{Effect of momentum cutoff on lattice formation. (a)
Evolution of the condensate energy for the unimposed case
(grey/brown line), where the allowable $\it k$-values are
restricted by the numerical grid to a {\it square} region
$[k_x,k_y\leq k_{\rm max}]$, and imposed {\it circular} regions
with radius $k_{\rm cut}=k_{\rm max}$ (solid black line),
$0.8k_{\rm max}$ (dashed black line), and $0.6k_{\rm max}$ (dotted
 black line).  Here $k_{\rm max}=2\pi/\delta$ is the maximum allowable ${\it
k}$-value, which is set by the isotropic grid size $\delta$. (b)
The norm of the condensate for the same cases. (c) Number of
vortices in the condensate for these four cases.} \label{mom}
\end{center}
\end{figure}

The numerical grid on which the simulations are performed
naturally imposes a momentum cutoff into the system.  In our
square 2D system, with isotropic grid spacing $\delta$, the $\it
k$-values are restricted to a square region $|k_{x,y}|\leq k_{\rm
max}$, where $k_{\rm max}=2\pi/\delta$. We employ an isotropic
grid spacing $\delta=0.25\xi$, which is sufficiently small that it
does not restrict directly the dynamics of the condensate.
However, it will affect the thermal component of the system.  High
energy atoms will `evaporate' from the system when they exceed
this cutoff.  For example, consider the total energy and norm
during the lattice formation process (figure \ref{mom}, solid
black lines) . The total norm does not visibly decrease, yet a
very small proportion of thermal atoms are lost. These are the
most energetic atoms, causing a gradual decrease in the energy at
later times (figure \ref{mom}(a), solid black line).

We investigate the role of the classical thermal cloud in the
lattice formation process by varying the cutoff in {\it k}-space.
To impose a cutoff at $k_{\rm cut}$, we Fourier transform into
{\it k}-space, set $|\psi(k_x,k_y)|=0$ for values which satisfy
$(k_x^{2}+k_y^{2})^{1/2}>k_{\rm cut}$, and then inverse Fourier
transform to obtain the modified spatial wavefunction.  This is
repeated regularly throughout the simulation (every 100 time
steps).  Note that the imposed cutoff is circular in {\it
k}-space, in contrast to the square cutoff induced by the grid.

Figure \ref{mom} also shows the dynamics for various imposed
circular cutoffs: $k_{\rm cut}=k_{\rm max}$ (grey/brown solid
line), $0.8 k_{\rm max}$ (grey/brown dashed line), and $0.6k_{\rm
max}$ (grey/brown dotted line). These cases exhibit some common
features. Despite the reduction in the population of thermal
atoms, a vortex lattice forms in all cases.  Importantly, the
lattices contains approximately the same number of vortices
(figure \ref{mom}(c)). The peak condensate energy (figure
\ref{mom}(a)) is slightly lower than the case of no imposed cutoff
(black line). In the crystallisation phase the energy decreases
exponentially, and at a faster rate than the case of no imposed
cutoff. Similarly, for the norm (figure \ref{mom}(b)), atoms are
lost during the lattice formation process, and the rate of loss
increases as the cutoff is reduced. These features are to be
expected since lowering the cutoff `evaporates' more energetic
atoms from the system. However, the number of vortices in the
condensate does not visibly decrease, indicating that this
energy/particle loss has little or no affect on the vortex
dynamics in the central condensate.

What is somewhat surprising is the significant difference in the
energy curves between the square cutoff (figure \ref{mom}(a),
black solid line) and the circular cutoff with the same width
(figure \ref{mom}(a), grey/brown solid line). Clearly a large
amount of energy must acculumate in the high-{\it k} corners of
{\it k}-space.

For the imposed cutoffs, the energy has very similar dynamics,
despite the large variations in the cutoff. In particular, the
exponential decay of the energy during the crystallisation stage
occurs on similar timescales. Furthermore, the number of vortices
in the final lattice is insensitive to the cutoff.  This suggests
that the formation of the lattice is independent of the number of
thermal atoms, and therefore independent of temperature.
Experimental evidence to suggest that the crystallisation stage is
temperature-independent has been presented in
\cite{hodby,abo-shaeer2002}.

\section{Effect of terminating the rotation}
\label{sec:term} So far we have considered a condensate in a trap
which is under {\it continuous} rotation.  Here we study the
effect of terminating the rotation after some time. We first
consider the case of the stable quadrupole mode (described in
section \ref{sec:quad}). This cycle is driven and maintained by
the continuous trap rotation. Terminating the rotation freezes the
energy and elongation of the condensate. The quadrupole mode
ceases although the elongated condensate stills rotates at the
rotation frequency.

\begin{figure}[h]
\begin{center}
\includegraphics[width=4.5cm,angle=-90]{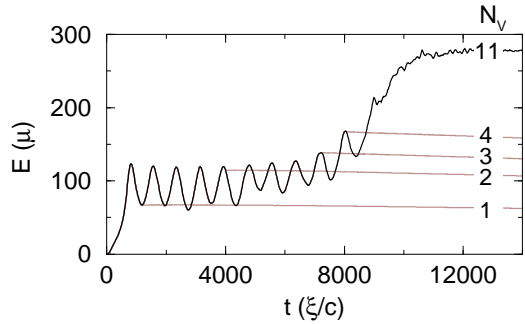}\\
\caption{Effect of terminating the rotation. Evolution of the
condensate energy for continuous rotation (black line), and
rotation which is terminated at times $t=$ (i) $1180$, (ii)
$4000$, (iii) $7200$, and (iv) $8000~(\xi/c)$.  The number of
vortices produced in each case is indicated.} \label{termination}
\end{center}
\end{figure}

Figure \ref{termination} presents results for the unstable
quadrupole regime. Here the energy for the case of continuous
rotation (black line) is compared with cases where the rotation is
turned off at various times (grey/brown lines). Once the rotation
is terminated, no more energy can be coupled into the system, and
the total energy becomes approximately fixed (some energy is lost
due to the finite momentum cutoff - see section
\ref{sec:thermal}). In all cases shown a vortex lattice is
ultimately produced, including the case where the rotation is
terminated after just one quadrupole oscillation.  This implies
that once the instability has been started, i.e. the fragmentation
during the first quadrupole oscillation, the full vortex formation
process will occur.

An interesting effect of terminating the rotation is that, because
the energy of the system becomes fixed, it allows control over the
number of vortices in the final lattice.  For example, under
continuous rotation the final state in figure \ref{termination}
(black line) contains a maximum value of 11 vortices. However, for
the cases where the rotation is terminated (grey/brown lines) the
final state contains $1,2,3$ and $4$ vortices. In principle using
this technique it should be possible to generate final lattices
containing any number of vortices up to the maximum value.

\section{Critical rotation}
\label{sec:crit}

The fast rotation of condensates at the critical frequency
$\Omega=\omega_r$ and above is of current interest, with the
prospect of ultimately generating exotic phenomena such as vortex
melting and a quantum Hall state of bosons \cite{quantum_hall}.
For a gas of classical non-interacting particles under critical
rotation the centrifugal force equals the trapping force, and the
particles are no longer confined \cite{butts}. Super-critical
rotation has been studied experimentally in a harmonic trap
featuring an additional quartic term to avoid this singularity
\cite{fast_rot}. Rosenbusch {\it et al.} \cite{crit_rot}
investigated critical rotation in a rotating trap both
experimentally and theoretically, and showed that, while the
non-interacting (non-condensed) gas explodes, repulsive
interactions can hold the condensate together. This occurs if the
trap ellipticity does not exceed a critical value (of the order of
$\varepsilon\sim0.1$). Then the centre of mass of the condensate
is destabilised and spirals away from the trap centre.  However,
if $\varepsilon$ exceeds the critical value, the condensate also
undergoes explosive dynamics. Note that the values of ellipticity
used throughout this paper are well within this critical
ellipticity.

\begin{figure}[b]
\begin{center}
\includegraphics[width=13cm,angle=0]{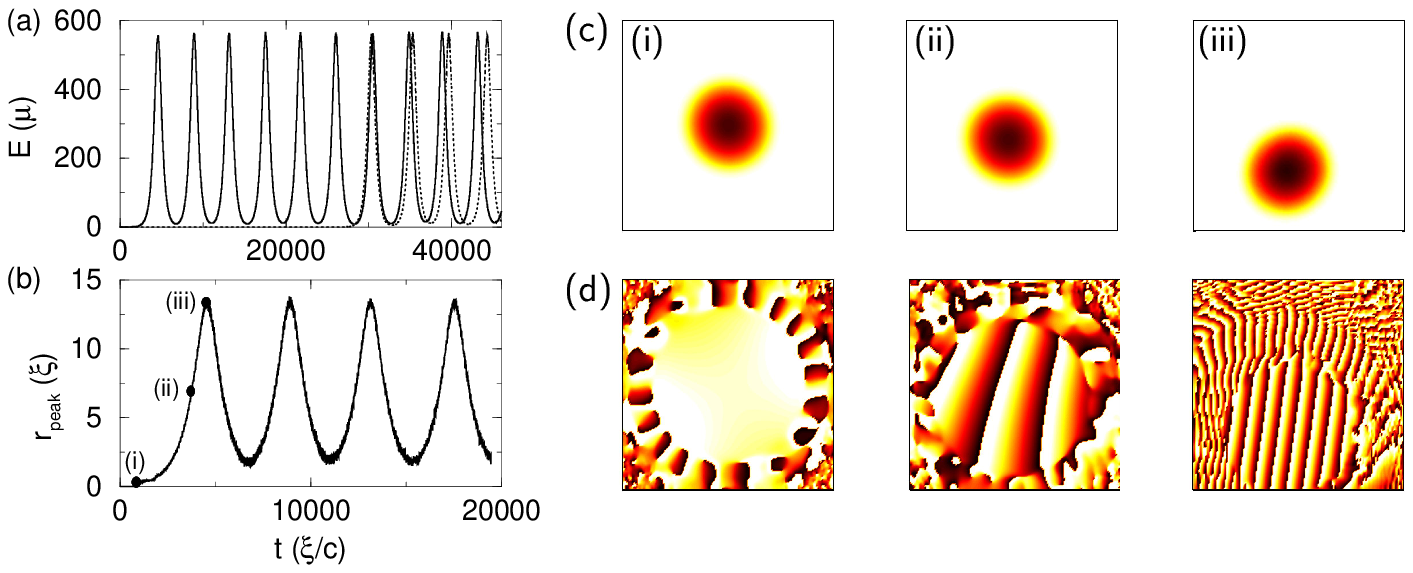}
\caption{Critical rotation. (a) Evolution of the total condensate
energy for critical rotation $\Omega=\omega_r$ with trap jitter
$\gamma=0.1$ (solid line line), and with no trap jitter (dotted
line). (b) Radius of the condensate peak (with trap jitter). (c)
Density and (d) phase plots of the condensate during the
centre-of-mass precession at times $t=$ (i) $500$, (ii) $3500$,
and (iii) $4500~(\xi/c)$. These times are indicated in (b).}
\label{com}
\end{center}

\end{figure}

We now consider the response of our system to critical and
super-critical trap rotation. As shown in figure \ref{quad_mode},
there is no strange behaviour in the quadrupole mode as the
critical point is reached and exceeded - the frequency/amplitude
continue to increase/decrease in a smooth manner.  However, we do
observe additional dynamics under critical rotation, namely the
periodic spiralling of the condensate in and out of the trap
centre.  This is in partial agreement with Rosenbusch {\it et al.}
\cite{crit_rot}, who conclude that spiralling condensate does not
return to the centre of the trap. As with lattice formation, these
dynamics require the initial two-fold symmetry of the system must
be broken.  Although this will occur slowly due to numerical
noise, we impose a trap jitter to speed up the dynamics, unless
stated otherwise. Figure \ref{com}(a)(solid line) shows the
evolution of the condensate energy for critical rotation
$\Omega=\omega_r$. The energy undergoes huge oscillations. This
corresponds to the periodic centre-of-mass motion of the
condensate to large radii in the trap (figure \ref{com}(b)).
Quadrupolar shape oscillations of the condensate still occur, but
the energy amplitude of the quadrupole oscillations (see figure
\ref{quad_mode}(b)) is three orders of magnitude smaller than
these oscillations. In the absence of an imposed trap jitter
(dotted line in figure \ref{com}(a)), the centre-of-mass dynamics
begin after the system has evolved for some time at low energies.

Figure \ref{com}(b)-(c) shows the density and phase during the
centre-of-mass oscillations. The condensate begins with
approximately circular shape and uniform phase (snapshot (i) in
figure \ref{com}(c)-(d)). It slowly drifts away from the trap
centre (figure \ref{com}(c), snapshot (ii)) while precessing
quickly at the rotation frequency. A striking quasi-linear phase
structure develops (figure \ref{com}(d), snapshot (ii)), which
represents the bulk motion of the condensate in the tangential
direction.  At a later time (figure \ref{com}(c)-(d), snapshot
(iii)), the condensate is further from the trap centre, and here
the linear phase structure is more concentrated.  Periodically the
condensate returns back to the trap centre.

\begin{figure}[t]
\begin{center}
\includegraphics[width=4.cm,angle=-90]{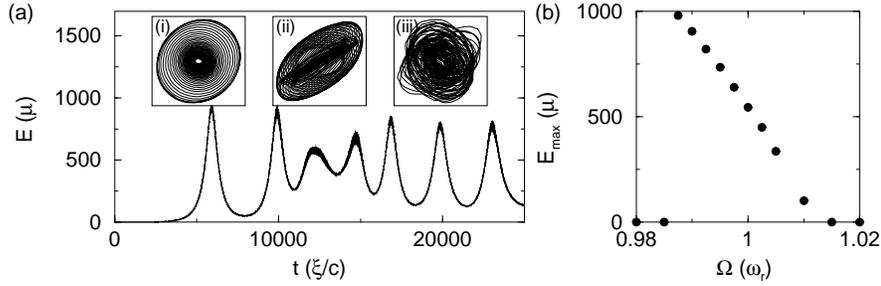}
\caption{Rotation at $\Omega=0.99\omega_r$. (a) Condensate energy
for a rotation frequency $\Omega=0.99\omega_r$ (with trap jitter
$\gamma=0.1\xi$). Insets show the path of the condensate peak
during the times (i) $3000-6000$, (ii) $9000-12000$, and (iii)
$18000-23000~(\xi/c)$. (b) Initial energy amplitude $E_{\rm max}$
of the centre-of-mass motion.}\label{com2}
\end{center}

\end{figure}

We observe that this centre-of-mass motion occurs in a range of
frequencies around the critical frequency.  However, the
centre-of-mass precession is not always of constant amplitude and
frequency. Figure \ref{com2}(a) shows the condensate dynamics for
a rotation frequency of $\Omega=0.99\omega_r$. Here the energy
oscillates with varying frequency and amplitude. As shown in the
insets of figure \ref{com2}(a), this is due to the condensate
moving in an ellipse, rather than a circle, with the ellipticity
and axes of the ellipse clearly changing with time.

We characterise the size of the centre-of-mass motion in terms of
the amplitude of the initial energy oscillations. Figure
\ref{com2}(b) shows this energy amplitude $E_{\rm max}$ for
rotation frequencies around the critical point.  We clearly see
that the centre-of-mass motion occurs in the narrow frequency
range $0.985<\Omega/\omega_r<1.015$.  In \cite{crit_rot} this
range is predicted to be
$\sqrt{1-\varepsilon}<\Omega/\omega_r<\sqrt{1+\varepsilon}$,
giving $0.987<\Omega/\omega_r<1.012$, which is in good agreement
with our observations. The irregular centre-of-mass motion occurs
in the lower part of this range, for $\Omega/\omega_r<0.9975$,
where the condensate is excited to the highest energies.

Above this critical region, the centre-of-mass oscillations cease
and the dominant excitation in the condensate returns to being the
quadrupole mode, albeit with very low amplitude (see figure
\ref{quad_mode}).

We have also considered the critical rotation of a condensate
which contains vortices, both by forming a vortex lattice at
$\Omega \sim 0.7\omega_r$ and suddenly switching the rotation to
$\Omega=\omega_r$, and by slowly ramping the rotation, through the
unstable regime, up to $\Omega=\omega_r$. In both cases, we
observe the condensate to undergo centre-of-mass oscillations in
the trap. However, here the oscillating condensate typically has a
diffuse and disordered appearance, and contains several vortices.

\section{Conclusions}
\label{sec:conc} In this paper, we have investigated the dynamics
of an atomic Bose-Einstein condensate in a weakly-elliptical
rotating trap using the two-dimensional Gross-Pitaevskii equation.
This was performed over a wide range of rotation frequencies
relevant to current experiments. The quadrupole mode of the
condensate plays a key role in the dynamics, with the fate of the
system depending crucially on the rotation frequency.  We resolve
the stable and unstable regimes of the quadrupole oscillations.
The unstable regime occurs around the resonant frequency of the
quadrupole mode, $\omega_{\rm Q}^0=\omega_r/\sqrt{2}$, and
rotation within this range ultimately leads to vortex lattice
formation.  Indeed, over the large range of rotations investigated
here, we only observe vortex nucleation to occur within this
range, i.e. vortex nucleation in elliptical traps only occurs
through the dynamical instability of the quadrupole mode. We
compare with lattice experiments and find good qualitative, and
sometimes quantitative, agreement, despite the limitations of the
model we employ (two-dimensional and zero-temperature). This
reinforces the idea that lattice formation is at heart a
two-dimensional effect. Furthermore, we find that the process is
insensitive to the cutoff of `thermal' atoms in the system,
suggesting that lattice formation is a zero-temperature
phenomenon.

By terminating the trap rotation during the lattice formation
process we show that it is possible to control the number of
vortices in the final lattice. Finally, we considered the response
of the system to critical rotation at the trap frequency, and
observe large centre-of-mass oscillations of the condensate,
without the explosive dynamics expected for a classical particle.

\ack We thank S. A. Gardiner for discussions and C. J. Foot for
the use of the experimental data in figure \ref{hodby}. We
acknowledge funding from the UK EPSRC.

\end{document}